\documentclass[english,letterpaper,twocolumn,showpacs,prl,aps]{revtex4}
\usepackage{graphicx}
\usepackage{amssymb}

\makeatletter

\large  

\input epsf

\makeatother

\usepackage{babel}
\makeatother

\begin{document}

\title{Optimal number of terms in QED series}

\author{Eugene B. Kolomeisky}

\affiliation
{Department of Physics, University of Virginia, P. O. Box 400714,
Charlottesville, Virginia 22904-4714, USA}

\begin{abstract}
In 1952 Dyson put forward a simple and powerful argument indicating that the perturbative expansions of QED are asymptotic.  His argument can be related to Chandrasekhar's limit on the mass of a star for stability against gravitational collapse.  Combining these two arguments we estimate the optimal number of terms of the QED series to be $3.1(137)^{3/2}\approx5000$. 

\end{abstract}


\maketitle

Quantum electrodynamics calculates measurable quantities through a power series that expands in powers of the fine structure constant $\alpha=e^{2}/\hbar c$.  Over sixty years ago Dyson \cite{Dyson} gave an argument indicating that QED series are asymptotic (i.e. there exists an optimal number of terms that best approximates the physical quantity being computed; keeping more terms does not improve but rather worsens the accuracy).  Dyson argued that this optimal number  is of the order $1/\alpha\approx137\gg 1$ which makes it clear why in practice the divergence he identified does not limit the utility of the perturbation theory.  Migdal and Krainov \cite{MK} improved on Dyson's estimate and arrived at the optimal number of terms of the order $1/\alpha^{3/2}\approx 137^{3/2}$.  The goal of this note is to show how the Migdal-Krainov estimate can be made more precise.  

In order to demonstrate that physical quantities regarded as functions of $\alpha$ or equivalently $e^{2}$ have a singularity at $e^{2}=0$ Dyson argued by contradiction.  If we assume that there is no singularity for $e^{2}=0$, the value of any physical quantity at $e^{2}=0$ should not depend on the way $e^{2}$ approaches zero -- whether it is from the $e^{2}>0$ or $e^{2}<0$ domains.  For $e^{2}>0$ the QED vacuum as a particle-free ground state is clearly stable.   On the other hand, if $e^{2}<0$ the particle-free vacuum is unstable and thus does not represent the lowest energy state of the system.  To clarify the meaning of the vacuum instability, Dyson interpreted the $e^{2}<0$ situation as describing a fictitious world where like charges attract and unlike charges repel.  In such a world were a quantum fluctuation to create $N$ electron-positron pairs, the electrons and positrons would spatially separate while the particles of the same type would segregate.  For $N\gg1$ the total energy of the well separated electron and positron regions will become lower than the energy of the particle-free vacuum because the energy penalty in creating pairs (proportional to their number $N$), will be inevitably compensated by the energy gain of attraction between like particles (proportional to the number of pair interactions $N(N-1)\approx N^{2}$).  Thus there is a smallest number of pairs $N_c$ beyond which the vacuum is unstable with respect to creation of another pair.  Since the vacuum is stable in the real $e^{2}>0$ world and unstable in the fictitious $e^{2}<0$ world, $e^{2}=0$ must be a singular point.  Creation of $N$ electron-positron pairs in the $e^{2}<0$ world corresponds to the $N$th order of the perturbation theory in the physical world, suggesting that the critical number of pairs $N_{c}$ corresponds to the optimal number of terms of asymptotic QED series.

Even though the instability of the $e^{2}<0$ vacuum has its origin in Coulomb attraction of like charges overcoming the rest energy of pairs ($2Nmc^{2}$), the critical pair number $N_{c}$ does not depend on the electron mass $m$.  This a consequence of the quantum nature of the problem (which puts a lower bound on the energy of a bound pair), and can be readily inferred from dimensional analysis:

The problem is fully specified by the dimensionless parameters $N$, $\alpha$, and by the electron Compton wavelength $\lambda=\hbar/mc$.  If there exists a critical number of pairs $N_{c}$, it can only be a function of the remaining independent dimensionless parameters of the problem.  However there is only one length scale $\lambda$ available, so that no dimensionless quantity can depend on it.  Therefore $N_{c}$ cannot depend on $\lambda$, and thus is independent of the electron mass $m$;  the only possible outcome is $N_{c}=f(\alpha)$, where $f$ is a yet unknown function which diverges in the non-interacting $\alpha\rightarrow 0$ limit.  

Dyson argued that a state whose energy is lower than that of the particle-free $e^{2}<0$ vacuum can be constructed from well-separated regions of electrons and positrons "without using particularly small regions or high charge densities, so that the validity of the classical Coulomb potential is not in doubt" \cite{Dyson}.  However, relativistic treatment of the particle dynamics is necessary, so that the Hamiltonian describing one of the regions occupied by $N$ like particles will be chosen in the semi-relativistic form 
\begin{equation}
\label{Hamiltonian}
H=c\sum_{i=1}^{N}\sqrt{p_{i}^{2}+(mc)^{2}}-e^{2}\sum_{i<j}\frac{1}{|\textbf{r}_{i}-\textbf{r}_{j}|}
\end{equation}
i. e. the particles obey the relativistic dispersion relation but attract each other according to the classical non-relativistic Coulomb law.  The quantum ground state of the Hamiltonian (\ref{Hamiltonian}) is a result of the interplay between the degeneracy pressure of confined fermions that tends to expand the region and the $1/r$ attraction among them that causes contraction.  A mathematically identical situation is encountered in the study of the gravitational equilibrium of bodies of large mass \cite{LL5} where it is known that there is the Chandrasekhar-Landau (CL) mass limit, beyond which the degeneracy pressure can no longer prevent gravitational collapse. The CL mass limit directly translates into an expression for the optimal number of terms $N_{c}$ of the QED series as explained below.  The relevance of the semi-relativistic Hamiltonian of the form (\ref{Hamiltonian}) to the problem of gravitational collapse was realized by L\'evy-Leblond \cite{Leblond} whose analysis we now follow.      

If $p$ is an average momentum of the particle while $r$ is an average distance between the particles, in the $N\gg1$ limit the total energy of the interacting system described by the Hamiltonian (\ref{Hamiltonian}) can be estimated as 
\begin{equation}
\label{general_energy estimate}
E(N)\simeq Nc\sqrt{p^{2}+(mc)^{2}}-\frac{N^{2}e^{2}}{r}
\end{equation}
The Pauli principle requires that there not be more than one fermion per de Broglie wavelength $\hbar/p$.  Then $N$ particles will occupy the total volume of about $N(\hbar/p)^{3}$ and the average distance between the particles (and the size of the region) is of the order $r\simeq N^{1/3}\hbar/p$ which transforms the expression (\ref{general_energy estimate}) into 
\begin{equation}
\label{E_of_p}
E(N)\simeq Nmc^{2}\left (\sqrt{x^{2}+1}-\alpha N^{2/3} x\right ),~~~x=\frac{p}{mc}
\end{equation}
Since the energy per particle depends on the particle number $N$ through the $\alpha N^{2/3}$ combination and the critical number $N_{c}$ can only depend on $\alpha$, this gives the estimate of Migdal and Krainov, $N_{c}\simeq \alpha^{-3/2}$, whose origin is in the fermion nature of the particles.  

On the other hand, if the latter were bosons, the average distance between the particles would be dictated by the uncertainty principle, $r\simeq\hbar/p$.  Substituting this into Eq.(\ref{general_energy estimate}) would give an expression similar to (\ref{E_of_p}) except that $\alpha N^{2/3}$ would be replaced by $\alpha N$.  This reproduces Dyson's estimate $N_{c}\simeq \alpha^{-1}$ whose origin is in the bosonic nature of the particles.

As a function of the free parameter $x$ the function (\ref{E_of_p}) has a minimum for $N<N_{c}\simeq \alpha^{-3/2}$ while for $N>N_{c}$ the minimum does not exist and the energy can be made as negative as needed by choosing large enough $x$:  the Hamiltonian is unbounded from below and the system faces inevitable collapse analogous to the gravitational collapse \cite{LL5}.  Indeed, whenever the minimum of (\ref{E_of_p}) exists, the lowest energy of the two well-separated electron-positron regions and the size of each region $r_{0}$ can be estimated as
\begin{equation}
\label{lowest_energy}
2E_{0}\simeq2Nmc^{2}\left [1-\left (\frac{N}{N_{c}}\right )^{4/3}\right ]^{1/2}
\end{equation} 
and 
\begin{equation}
\label{size}
r_{0}\simeq \lambda N_{c}^{2/3}N^{-1/3}\left [1-\left (\frac{N}{N_{c}}\right )^{4/3}\right ]^{1/2}
\end{equation}
respectively.  Since the energy (\ref{lowest_energy}) is counted with respect to that of the particle-free vacuum, the states with $N<N_{c}$ have higher energy than that of the original vacuum and the latter is stable.  Therefore creation of the electron-positron pairs is not energetically favorable until $N$ reaches $N_{c}$.  This is when both $2E_{0}$ and $r_{0}$ vanish with the latter being analogous to gravitational collapse \cite{LL5}.  Specifically the dependence of the size of the region on the number of particles it contains (\ref{size}) parallels the dependence of the radius of gravitating region on its mass \cite{LL5}.   The correspondence between the $N=N_{c}$ condition with that of gravitational collapse can be further coraborated by looking at the ultrarelativistic (or $m=0$) limit when the estimate (\ref{E_of_p}) becomes
\begin{equation}
\label{ultra}
E'(N) \simeq Ncp\left [1-\left (\frac{N}{N_{c}}\right )^{2/3}\right ]
\end{equation} 
We now see that $N=N_{c}$ corresponds to the state of neutral equilibrium when an arbitrary region size is allowed.  For $N>N_{c}$ the region would catastrophically contract with the energy decreasing without bound which corresponds to rapid disintegration of the vacuum.  Since the ultrarelativistic case is exactly solvable and $N_{c}$ is mass-independent, the value of $N_{c}$ follows from the expression for the CL mass limit \cite{LL5}
\begin{equation}
\label{CL_limit}
N_{c}=3.1\alpha^{-3/2}\approx 5000
\end{equation}
This improves on the Migdal-Krainov estimate of the optimal number of terms in QED series and demonstrates once again that asymptotic nature of QED series is only of purely theoretical interest.

The author is grateful to P. Arnold and J. P. Straley for valuable comments.  This work was supported by US AFOSR Grant No. FA9550-11-1-0297.

\end{document}